\documentclass[%
 aip,
 amsmath,amssymb,
 reprint,%
]{revtex4-1}

\usepackage{graphicx}
\usepackage{dcolumn}
\usepackage{bm}

\usepackage[utf8]{inputenc}
\usepackage[T1]{fontenc}
\usepackage{mathptmx}
\usepackage{etoolbox}

\makeatletter
\def\@email#1#2{%
 \endgroup
 \patchcmd{\titleblock@produce}
  {\frontmatter@RRAPformat}
  {\frontmatter@RRAPformat{\produce@RRAP{*#1\href{mailto:#2}{#2}}}\frontmatter@RRAPformat}
  {}{}
}%
\makeatother
\begin{document}

\title{Anisotropic plates with architected tendon network}
\author{Md Shahjahan Hossain}
 \affiliation{
Department of Mechanical and Aerospace Engineering, University of Central Florida. Orlando, FL
}%
\author{Hossein Ebrahimi}%
\affiliation{ 
Department of Mechanical and Aerospace Engineering, University of Central Florida. Orlando, FL
}%

\author{Ranajay Ghosh}
\email{ranajay.ghosh@ucf.edu}
\affiliation{
Department of Mechanical and Aerospace Engineering, University of Central Florida. Orlando, FL
}%

\date{\today}

\begin{abstract}
We synthesize geometrically tailorable anisotropic plates by combining button shaped fish-scale like features on soft substrates, then lacing them with high-stiffness strings. This creates a new type of biomimetic architectured structure with  multiple broken symmetries. First, the tendons and scales together break the symmetry of the bending response between the concave and convex curvature. Next, the weave pattern of the tendons further breaks symmetry along the two directors of plates. The anisotropy is clearly evident in 3-point bending experiments. Motivated by these experiments and the need for design, we formulate the analytical energy-based model to quantify the anisotropic elasticity and tailorable Poisson's ratio. The derived architecture-property relationships can be used to design architected tendon plates with desirable properties. 
\end{abstract}

\maketitle




Slender unremarkable substrates can be transformed into materials with extraordinary mechanical properties with an additional metastructure on one side. This has recently been exploited in biomimetic fish scale-inspired designs \cite{ebrahimi2021fish, buehler2006nature, wegst2015bioinspired, helfman2009diversity, ghosh2014contact, hossain2022fish}, membrane tensegrity \cite{skelton2001introduction, lu1998optimal, zhang2015tensegrity} and metasurfaces \cite{he2018high, arbabi2017planar, badawe2016true, martini2014metasurface}. However, adapting them for active structures require complex integration of stimuli-repsosive materials or intricate fluidic circuits. Thus, many of the benefits accrue with additional complexity.  Hence, tailorability of properties become increasingly difficult putting a limit on their ultimate technological impact. 

\begin{figure*}
\includegraphics[scale = 0.35]{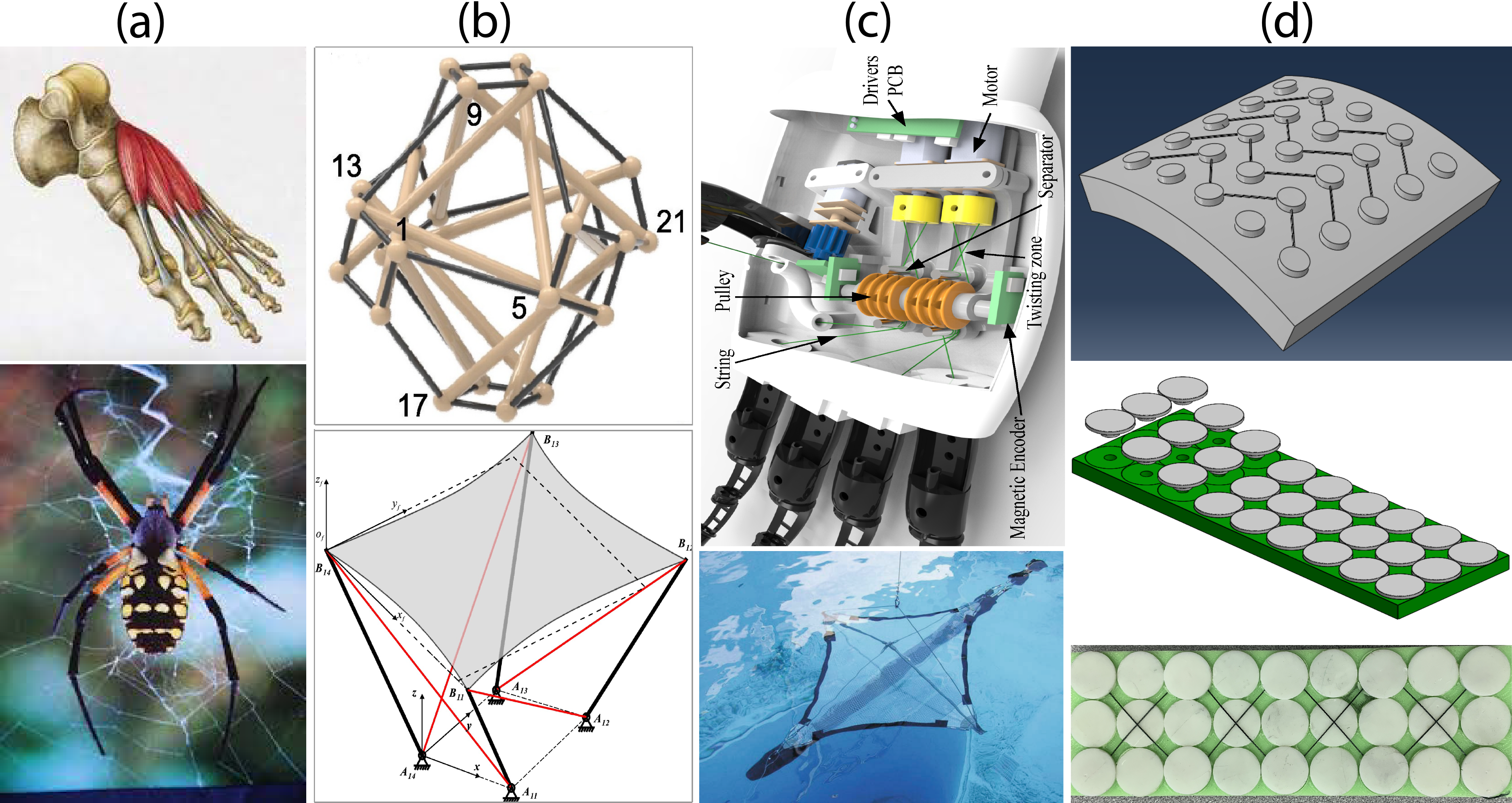}
\caption{\label{fig:1} String/tendon structures a) Sting/tendon structure in nature:  toe joints with tendon-bone structure (top) and spider fiber (often referred to as the golden
orb weaver) (bottom) (\cite{skelton2009tensegrity} b) two types of string system: Z-based truncated regular octahedral tensegrity system (top) and four-bar tensegrity-membrane system \cite{zhang2018automatically} and string membrane system \cite{yang2016modeling} c) bionic hand (top) \cite{tavakoli2015uc} and wave-powered station-keeping buoy (bottom) \cite{skelton2009tensegrity} d) illustration of String-substrate system used in this study.}
\end{figure*}

Tendons, which can be used to transmit force, Fig \ref{fig:1}(a) are string like structures whose intricate arrangements are often exploited in nature, for instance to transfer force from muscles to bones. Thus strings are one of the most versatile structure to act and distribute forces. They find numerous engineering applications such as tensegrity, used in tensegrity, Fig. \ref{fig:1}(b), lightweight deployable structures and soft robotics, Fig. \ref{fig:1}(c). Thus, strings can provide an easy way to impose tension and force on membranes without additional actuator complexity. Thus woven and tendon designs \cite{jeong2021reliability} remains one of the most exciting areas in modern soft and wearable applications. With the advent of new types of fibers\cite{bundhoo2009shape}, periodic tendon networks can create novel metastructures capable of tunable and tailorable mechanical properties.  The tendons themselves could be replaced by actuators such as coiled actuators \cite{zhang2020compliant, souza2019series}, thermal actuators \cite{rodrigue2017overview, knick2019high}, and humidity actuators \cite{wang2018natural, pu2022review}. Strings with suitable mechanical configurations, such as twisted strings of actuators and strings of pearls have advantages due to their lightweight, tailorability, and silent operation. \cite{rutherford2015effect, usman2017passive, souza2019series, jang2022active, popov2013bidirectional}.

Specifically, the architecture of the periodic network or weave of the tendons can be exploited to create desirable mechanical stiffness when integrated with a substrate. Such designs lead to metastructures with unique and tailorable mechanical properties not seen in conventional designs. In Fig. \ref{fig:1}(d) different tendon structures are shown that were used in our study. The string side bending (positive) can result in additional stiffness due to string tension whereas the other side retains the behavior of the underlying substrate only to jam/lock as the buttons touch each other. This creates a tailorable bi-directional metastructure.  In spite of such promises, little work has been done to systematically understand and model the resulting metastructure. The lack of architecture-property relationships hinders further design driven growth and exploration. Relationships between anisotropic and asymmetric  plate bending, effect on Poisson ratio and contrast in bi-directional properties are not in the literature. We address this gap for the first time by systematic derivation of closed form relationships using energy principles \cite{mousanezhad2016elastic}.

To motivate our study, we fabricated the tendon samples using a multi-step manufacturing method that included a soft substrate, a button (used as an anchor), and string. We create a mold with dimension $212mm(Lenght)\times 65mm  (Width)\times 10mm(Height)$. A 3D (Ultimaker\textsuperscript{TM}) printer was used  to fabricate the mold of the substrate and button (Polylactic acid, PLA)  ($E_{PLA}  \sim 3 GPa$). 
In order to create the soft substrate, mixtures of silicon rubber (Dragon Skin\textsuperscript{TM} 10) and curing material $(1:1)$ were cast into the mold. The button was encapsulated into the soft substrate without using adhesive after allowing the substrate to cure and which was then removed from the mold. Finally, we wound the string (Power Pro Spectra Fiber Braided Line) ($dia \sim 0.35mm$) in four different arrangements to conduct the experiment, the string were wrapped once around the buttons..

To demonstrate the difference between various weaves, a 3-point bending experiment was performed using MTS Insight\textsuperscript{\textregistered} under displacement control at a crosshead speed of 1mm/s with cross-head displacement from $0-50 mm$.  Here, we tested eight distinct samples with buttons placed next to each other with minimal gap. These samples included: without string ($S_0$), with one zigzag string ($Z_1$) (at an angle of $45^{\circ}$), with two zigzag string ($Z_2$) (at an angle of $45^{\circ}$), and two straight string (at an angle of $0^{\circ}$), as shown in table I\ref{Table:1}. We used string side loading (positive loading) and button side loading (negative loading) for obtaining the stiffness. The load cell values were acquired in relation to the cross-head displacement.

\newcolumntype{L}[1]{>{\raggedright\let\newline\\\arraybackslash\hspace{0pt}}m{#1}}
\begin{table}
\label{Table:1}
\centering
\begin{center}
\caption{Various weaves of string-substrate samples used in this work}
\begin{tabular}[t]{>{\centering}L{7em}|>{\centering} L{7em}| L{11em}}
\hline
Test Samples & Symbols & Representation of loading  \\
\hline
Without String & $S_0$ & 1. String side loading\\ 
\cline{1-2}
One ZigZag String & $Z_1$ & {\begin{minipage}{.2\textwidth}
      \includegraphics[width=\linewidth, height=17mm]{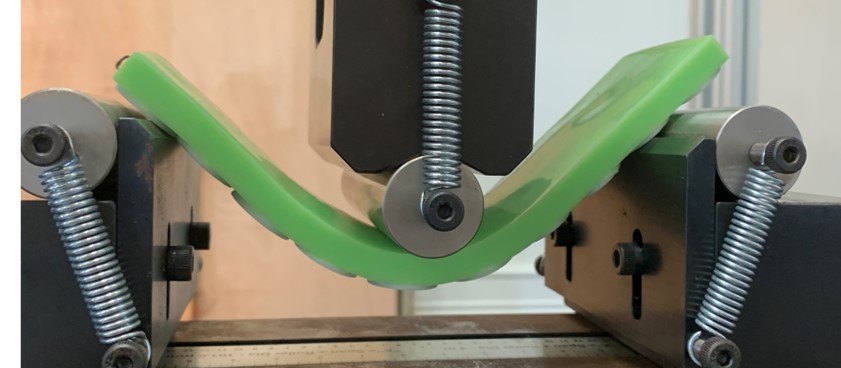}
    \end{minipage}}\\ 
\cline{1-2}
Two ZigZag String & $Z_2$ & 2. Button side loading \\
\cline{1-2}
Two straight String & $S_2$ & {\begin{minipage}{.2\textwidth}
      \includegraphics[width=\linewidth, height=17mm]{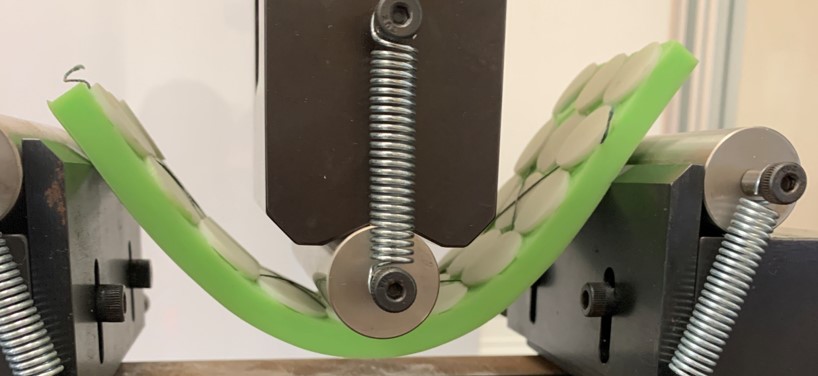}
    \end{minipage}} \\ 
\hline
\end{tabular}
\end{center}
\end{table}

\begin{figure}
\includegraphics[scale = 0.40]{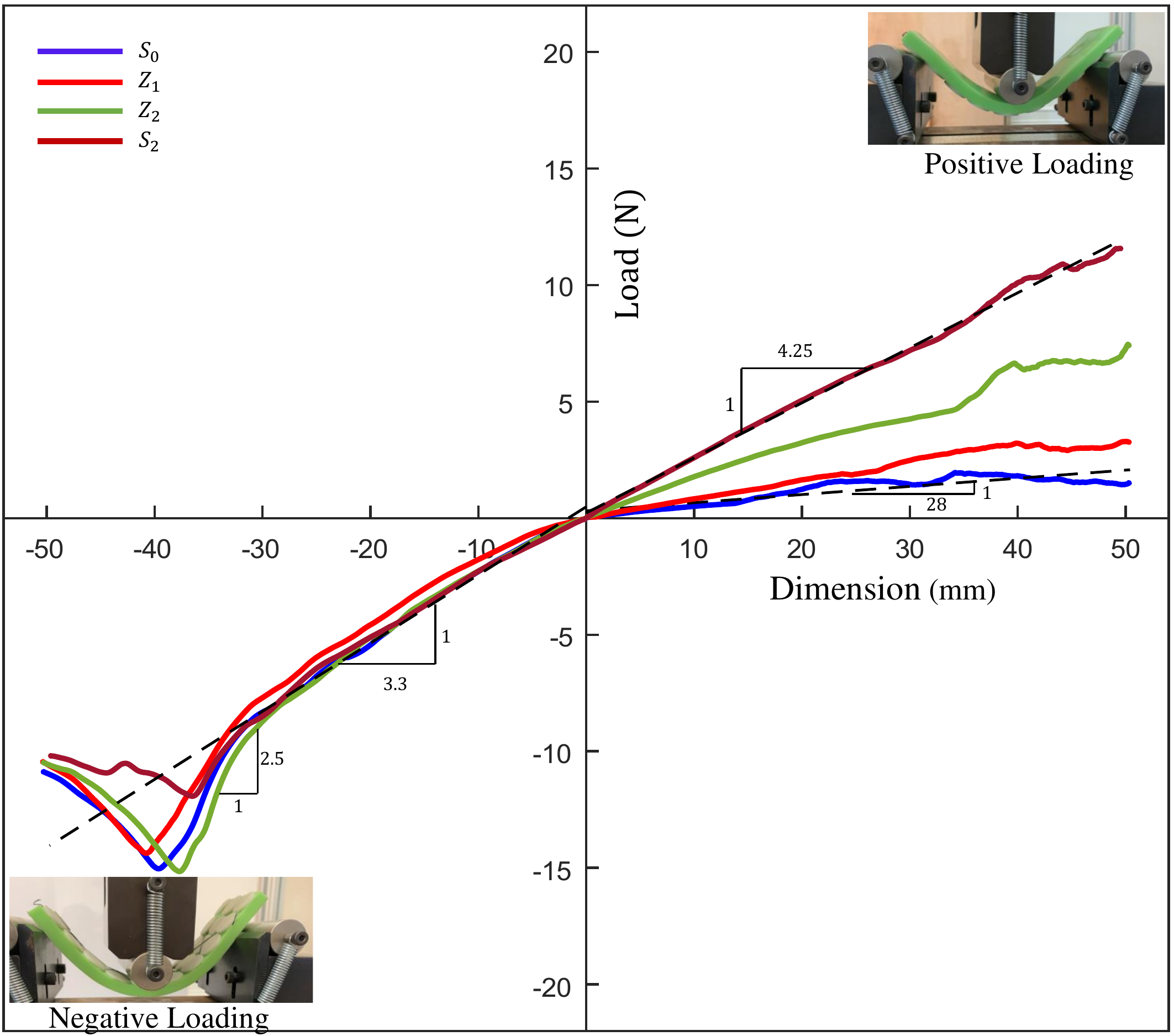}
\caption{\label{fig:4} Load vs displacement curve for 3 point bending of tendon network substrate on two sides of bending. The effect of weave and direction of bending have a distinct effect on bending stiffness. Positive side bending is intricately related to the type of weave.}
\end{figure}

The load vs displacement curve for positive and negative loading is shown in Fig. \ref{fig:4}.  The load-displacement curve shows two fundamentally different behaviors on the sides of loading. The string side stiffness is significantly higher for $Z_2$ and then decreases in the order of $Z_1$ and $S_2$ when compared to baseline (no string). Eventually, the stiffness plateaus due to string slippage and transition from string extension mode to button bending mode that show similar stiffness for all samples. On the other hand, the behavior on the reverse side is unfettered by the strings as they are not engaged. Eventually, the structure reaches a geometrically dictated locking/jamming configuration when the buttons touch each other. These results inspire us to develop a mathematical architecture-property for the string-substrate systems.


\begin{figure*}
\includegraphics[scale = 0.50]{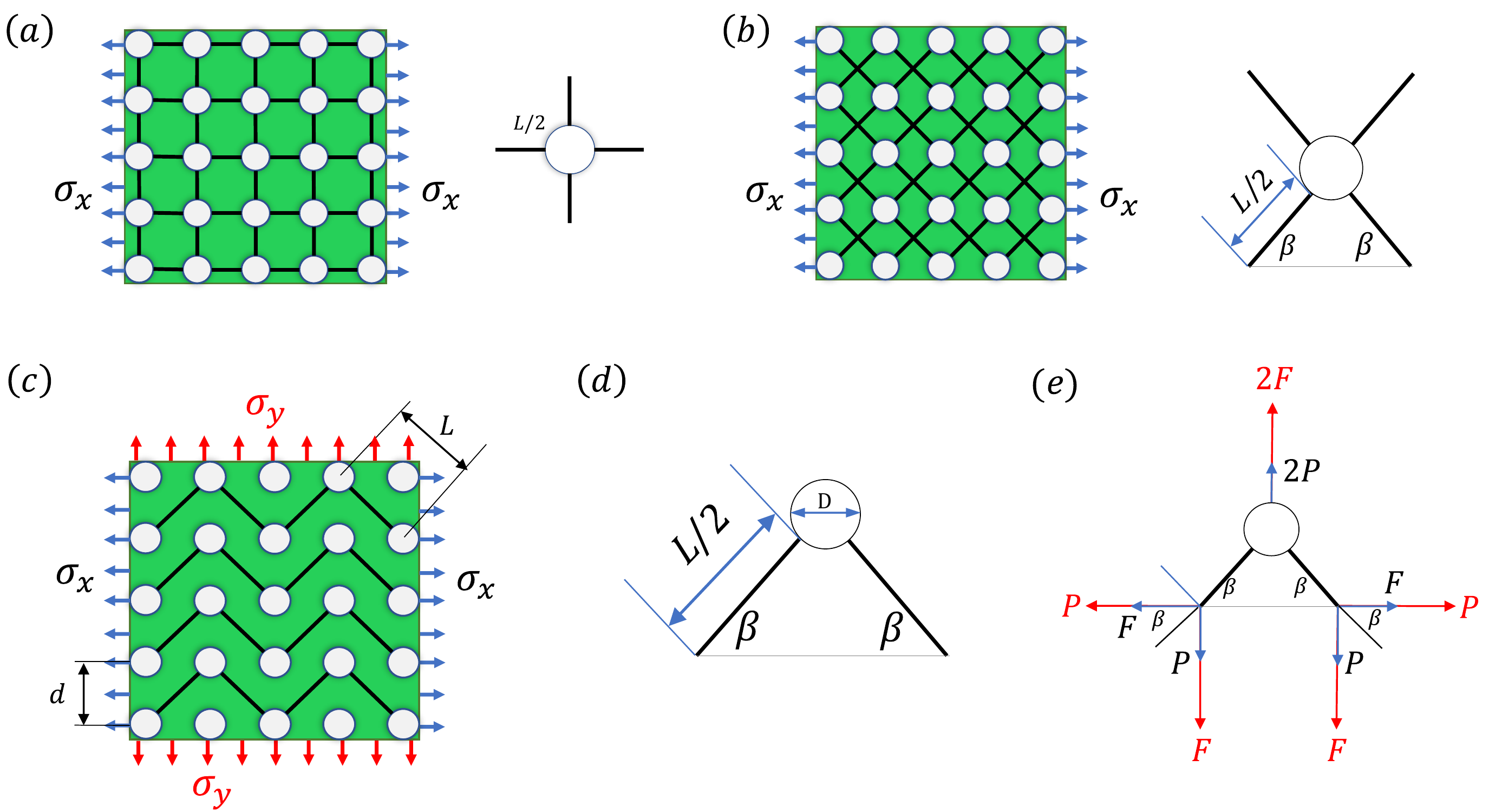}
\caption{\label{fig:2}Figures in the upper row show string configuration and unit cell of a) $S_2$ and b) $Z_2$. Figures in the lower row show the string configuration of $Z_1$ c)  having stress in the x-axis and y-axis, d) unit cell, and e) showing forces applied to the unit cell for both stresses}
\end{figure*}

Our mathematical model considers a single type of unit cell that of $Z_1$ type. $Z_1$'s geometrical parameters can encompass all the other unit cells considered in this work. Fig. \ref{fig:2}(a-c) displays the three different types of strings and their corresponding unit cells.
For the unit cell, we treat string and substrate as elastic material and the buttons as rigid.  We take $E_b$ and $E_s$ as the modulus of elasticity of substrate and string respectively.

The length of the string, its radius, and the perpendicular distance between the two button edges are denoted as $L$, $r$, and $d$ respectively. To load the system, a stress of $\sigma{_x}$ in the direction of the x-axis and a stress of $\sigma{_y}$ in $y$-direction is applied. Fig. \ref{fig:2}(e) illustrates the force operating in the $x$-direction (blue) and $y$-direction (red) as a result of the stress. The forces acting in the $x$ direction and $y$ direction respectively are $F=\sigma{_x}(Lcos\beta)$ and $F=\sigma{_y}(Lsin\beta)$, where$\beta$ is the angle of sting with horizontal line. Note that this load is based on the assumption that in-plane condition on the tendon metastructure is that plain stress.


To find out the structure's Young's modulus and Poisson ratio in the $x$ and $y$ direction, virtual force $P$ is applied in the $y$-direction (blue) for $x$-directional stress. Another virtual force of magnitude $P$ is applied on the $x$-direction to calculate the Young's modulus and Poisson ratio in $y$-direction. Note that the actual magnitude of the fictitious force is unimportant for application of energy methods. the $x$-direction (red) for $y$-directional load as shown in Fig. \ref{fig:2}(e). Eq. (\ref{equation:1}) represents the strain energy of the unit cell due to stress in the $x$-direction and Eq. (\ref{equation:4}) represents the strain energy due to stress in the $y$-direction, where A is the cross-sectional area of cell walls and $I$ is the second moment of the area of the string structure. 

\begin{equation}
U=\frac{2{(Fcos\beta+Psin\beta)}^2\frac{L}{2}}{2E_sA}
  +2\int_{0}^{\frac{L}{2}}\frac{(x{(Fsin\beta-Pcos\beta)})^2}{2E_sI} \,dx
\label{equation:1}
\end{equation}

\begin{equation}
U=\frac{2{(Fsin\beta+Pcos\beta)}^2\frac{L}{2}}{2E_sA}
  +2\int_{0}^{\frac{L}{2}}\frac{(y{(Fsin\beta-Pcos\beta)})^2}{2E_sI} \,dy\\
\label{equation:4}
\end{equation}

From here, using Castigliano's theorem (i.e. $\delta=\frac{\delta{U}}{\delta{F}}\textbar{_{P=0}}$), we calculated the structure's average strain ($\epsilon{_x}=\delta_x/Lcos\beta$ for the $x$-direction and $\epsilon{_y}=\delta_y/Lsin\beta$ for the $y$-direction) (see Supplementary Material). Thus, the structure's modulus of elasticity was determined by the ratio of average stress $\sigma_x/\epsilon_x$ for $x$-direction and $\sigma_y/\epsilon_y$ for $y$ direction. Poisson's ratio for the necessary load directions was established using the ratio of the change of displacement in transverse direction to the axial direction (see Supplementary Material). For the string-substrate structure, we have the Poisson effect for both the string metastructure and membrane material. To determine the overall modulus of elasticity we volume average the elastic constants. Thus, for finding the overall Poisson ratio of the structure, we employed a volumetric weighted ratio (i.e. in $x$-direction, $\nu_x = (V_b\nu_b+V_s\nu_{sx})/(V_b+V_s)$) that combines the Poisson's ratios of the individual components to allow calculating the overall Poisson's ratio of the substrate (see Supplementary Materials).

Based on the developed material properties, we calculated the equation for the bending moments $M_x$, and $M_y$. using the Kirchhoff plate theory shown in Eq. (\ref{equation:11}) and Eq. (\ref{equation:12}). Here, $C_f$ is the correction factor that arises due to additional stiffness of the button inserts (inclusion effect). Such inclusion factors can be computed using FE simulations. The flexural rigidity of a plate is given by $D_x = \frac{wE_xh^3}{12(1-\nu_x^2)}$ and $D_y = \frac{wE_yh^3}{12(1-\nu_y^2)}$. Where $\nu_x$ and $\nu_y$  represent the combined Poisson ratio (that we achieved from volumetric weighted average of Poisson ratio for substrate and string), and $E_x$ and $E_y$ is the combined modulus of elasticity (from volumetric weighted average of modulus of elasticity for substrate and string) in the $x$ and $y$-directions, respectively (see Supplemental material).

\begin{equation}
M_x = C_fD_x(\kappa+\nu_x\tau)
\label{equation:11}
\end{equation}

\begin{equation}
M_y = C_fD_y(\tau+\nu_y\kappa)
\label{equation:12}
\end{equation}

\begin{figure*}
\includegraphics[scale = 0.38]{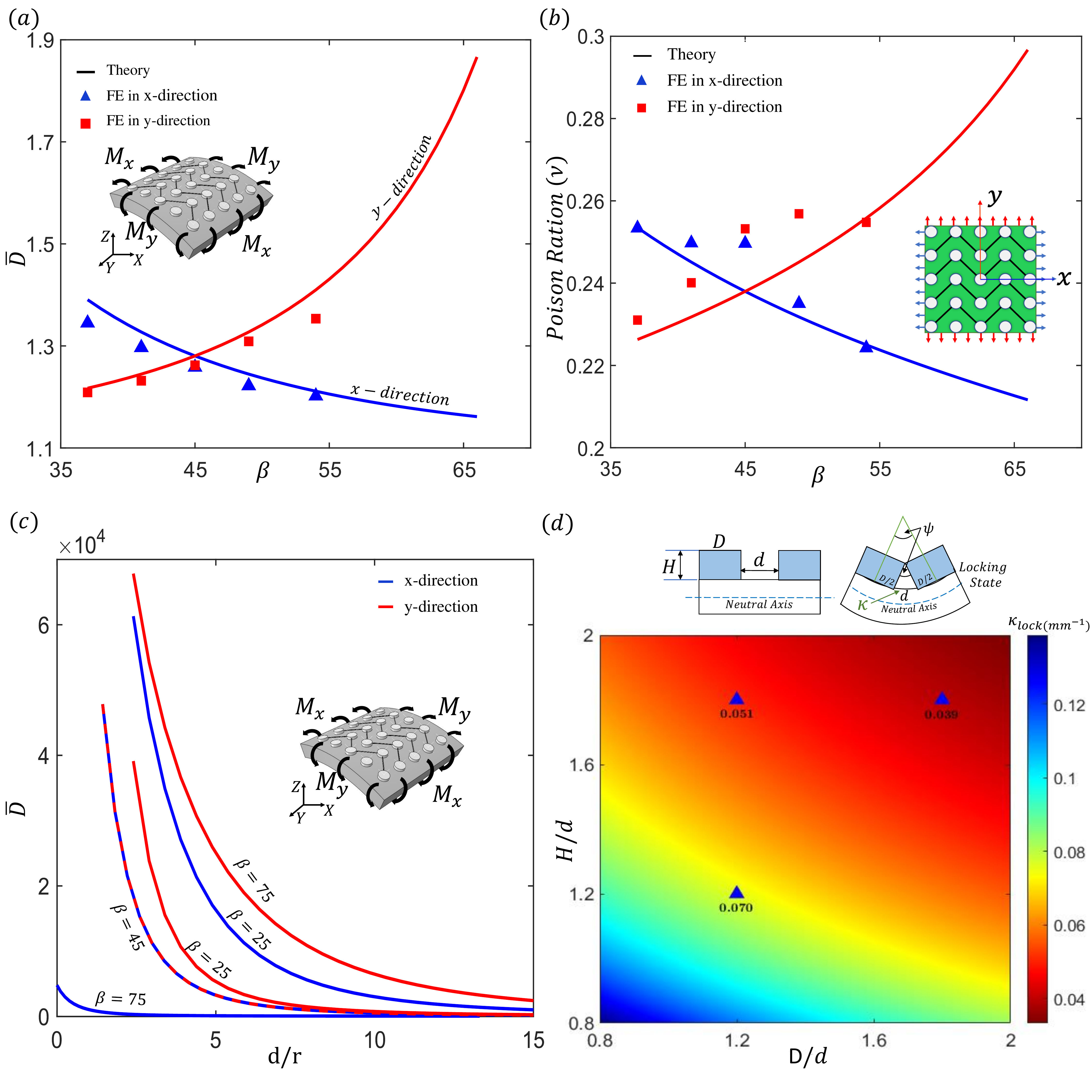}
\caption{\label{fig:5} a) Normalized bending stiffness $\bar{D}$ result for string-substrate systems due to the applied moment in $x$-direction and $y$-direction, b) Poisson ratio of string-substrate system ($\nu$) vs weave angle, for $x$ and $y$-axis directions, c) $\bar{D}$ with the change of $d/r$ at angle $25^\circ$, $45^\circ$, and $75^\circ$, d) unit cell geometry before and after locking (top) and phase plot of locking curvature with $H/d$ and $D/d$ ratios (bottom), where discrete plot denotes FE simulation results.}
\end{figure*}

In the negative loading, the strings are not engaged and hence the stiffness is independent of weave and the same as the substrate. However, eventually, at sufficiently large curvatures, the buttons engage with each other leading to a geometrically locked/jammed configuration, shown in Fig \ref{fig:5}(d). This locking curvature can be obtained via the geometry of the buttons and bending kinematics. For this purpose, we define $D$, $H$, $\psi$, and $d$ the button's diameter, the height of the button, substrate angular change (related to curvature $\kappa$), and the distance between the buttons respectively. From the Fig. \ref{fig:5}(d) (top) which represents a unit cell, we find two curvature relations $\psi_{lock}=d/H$ and $\psi_{lock}=\kappa_{lock}(D+d)$. This leads to button size and lattice spacing relation with curvature lock, $\kappa_{lock}=d/(d+D)H$.


For validating these results, we use commercially available finite element (FE) based simulation software ABAQUS/CAE 2021 (Dassault Syst\'emes). The digital models were made to conform to the analytical model. The buttons were subjected to rigid body constraints, which eliminated the need for button material attributes. During the simulation, the material properties have been used similar to the properties used for the calculation of mathematical modeling. The substrate's Poisson ratio and modulus elasticity were estimated to be $10 MPa$ and $0.2$, respectively. Also, the string's Poisson ratio and modulus elasticity were taken as $1100 MPa$ and $0.25$, respectively. These numbers are close to the physical system considered. However, the model is not tied to these exact numbers. A total of twelve unit cells were considered to represent the string-substrate system. The effect of non-linearity during the simulation was negligible and hence the bending of the structure was modeled using a static step with a linear geometry option. The center of the beam's cross-section was fixed, and rotation was applied to both ends of the $x$-direction and $y$-direction during the study of the respective directions. Sufficient mesh density was used in order to achieve mesh convergence. For the simulations, approximately 338868 Quadratic tetrahedron element C3D10 (standard quadratic 3D stress elements) were used that were standard quadratic 3D stress elements.

We first plot the bending rigidity of the tendon metastructures in both $x$ and $y$ direction according to the weave angle for positive direction bending. This is illustrated in Fig. \ref{fig:5}(a) which plots, nondimensionalized $\bar{D}$ (where, $\bar{D}=M/\kappa D_b$) against the angle for both the theoretical and FE outcomes. The FE models showed excellent match with analytical calculations. We found that with the increase of angle along the horizontal axis, the stiffness in the $x$-direction decreased, whereas the opposite is true for $y$-direction. Thus the contrast in anisotropy between the two directions can be increased sharply by changing the weave angle as the trends are opposite. Not surprisingly, at the $45^\circ$ angle of weave (balanced weave), the stiffnesses in both directions are the same. Hence, the overall structure becomes symmetric. It is worth noting that the change in stiffness in both directions with angle $\beta$ is nonlinear but with opposite rates of increase in two directions. The sensitivity to beta is much higher in $y$-direction at higher beta whereas the reverse is true for $x$-direction. Thus, much higher stiffness gains can be achieved in one direction with varying angle keeping the other direction stiffness relatively unchanged.

Fig. \ref{fig:5}(b) we plot the combined Poisson ratio of the string-substrate meta structure against different angles $\beta$ for both directions for positive direction bending. As the angle increases, there is a decrease in the effective poison ratio for the $x$-direction while the opposite holds true for the $y$-direction. We have once again observed a similar anisotropic trend in the Poisson ratio at different angles, as was previously seen in stiffness. The Poisson ratio shows some discrepancy between numerical and theoretical values. the average discrepancy was roughly 5\% and maximum 10\%. The major reason for the discrepancy could be amplification of errors arising from the two directions of calculations in the previous step. Another reason is the inaccuracies in measuring average Poisson's ratio from numerical calculations alone. We anticipate that larger number of unit cells would reduce this discrepancy. Similar to stiffness, the Poisson ratio also exhibits greater sensitivity in the $y$-direction as the angle $\beta$ increases.

Fig. \ref{fig:5}(c) represents bending rigidity with respect to normalized string length, i.e. $d/r$. In other words, the geometric tailorability of the substrate with respect to button lattice spacing. We find that bending rigidity significantly decrease for all angles with increasing button lattice spacing. The gains are nonlinear and the highest gains in rigidity occur at the lowest $d/r$. Interestingly, the loss in stiffness with increased lattice spacing can be counteracted by higher winding angle of the weave. These plots indicate a complex inter-relation between the string lattice spacing and string weave with respect to gain in bending stiffness. This would have implications in functionally graded designs with various string gages.

Fig. \ref{fig:5}(d) shows the phase plot of locking curvature at different $D/d$ vs $H/d$ in the negative direction. The locking curvature is a measure of range of motion of the structure in the negative direction. The plots show excellent agreement with FE results. 
The plot clearly shows that both wider and taller buttons reduce the range of motion. The \textit{isocurvature} lines appear to be hyperbolic. This is supported by the scaling law emerging from the locking relationship, $\kappa_{lock}=\frac{d}{d+D}\frac{1}{H}=\frac{1}{d}\frac{1}{1+D/d}\frac{1}{H/d}$. Thus, higher spacing  leads to rapid increase in range of motion. For very large spacing $d>>D$, we get $\kappa_{lock}\sim\frac{1}{H}$. This result is essentially same as the locking limit of scale covered substrates ~\cite{ghosh2014contact}. However, note that in practice this curvature may be larger than the physical limit of the substrate bending before it closes onto itself.


In conclusion in this paper, we derived architecture-property relationship. We motivate our work using the 3-point bending test on the samples revealed that the bending stiffness was a function of both the weave geometry and the spacing of the buttons over which winding was carried out. We derived structure-property relationships for the tendon network that allowed us to quantify the effect of weave parameter on stiffness. Our results revealed significant anisotropy of the resultant metastructure. The nature of the string allowed bi-directional asymmetry (between two directions of bending). The structure-property relationships were nonlinear with disproportionate sensitivity towards wound angles for some directions. Our calculations on Poisson's ratio also showed that it can be depressed by adjusting the weave. In spite of this auxeticity was not observed. 

Thus, our work lays the foundation of a new type of metastructure that is capable of building a system with tuneable material properties by utilizing the mathematical model established in this letter. Which could be used to make metamaterials. This string substrate system can also be used in soft robotics by swapping out the string for shape memory alloy. 

This work was supported by the United States National Science Foundation's Civil, Mechanical, and Manufacturing Innovation, CAREER Award \#1943886.

\appendix




\vspace{5mm}
\nocite{*}
\bibliography{String_paper.bib}

\end{document}



\title{Online Supplementary to Anisotropic plates with architected tendon network}
\author{Md Shahjahan Hossain}
 \affiliation{
Department of Mechanical and Aerospace Engineering, University of Central Florida. Orlando, FL
}%
\author{Hossein Ebrahimi}%
\affiliation{ 
Department of Mechanical and Aerospace Engineering, University of Central Florida. Orlando, FL
}%

\author{Ranajay Ghosh}
\email{ranajay.ghosh@ucf.edu}
\affiliation{
Department of Mechanical and Aerospace Engineering, University of Central Florida. Orlando, FL
}%
\date{\today}

\begin{abstract}
 
\end{abstract}

\maketitle


\section{Stress in $X$-direction}

To determine the material properties of the string-substrate structure due to applied stress mentioned in manuscript, we applied the Castigliano theorem ($\delta=\frac{\delta{U}}{\delta{F}}$) to the energy Eq. (\ref{equation:1}) for the $x$-direction, which provides the displacement in axial direction, $\delta{_x}= \frac{\delta{u}}{\delta{F}}\textbar{_{P=0}} = \frac{FL}{E_sA}(cos^2\beta)+\frac{FL^3}{12E_sI}(sin^2\beta)$. Following that, the structure's average strain $\epsilon{_x}$ in the $x$-direction was determined using $\epsilon{_x} = \frac{\delta{_x}}{Lcos\beta}$. The effective Young's modulus $E_{xu}$ of the structure is calculated as the ratio of average strain $\sigma{_x}$ to average strain ($\epsilon{_x}$). That provides, $E_{xu} = \frac{12AI}{12ILcos^2\beta+L^3Asin^2\beta}\times{E_s}$, here $E_{xu}$ is the elastic modulus of a unit cell. Finally, we multiplied it by the number of unit cells involved ($n_u$) and width ($w$) of the structure to find out the total modulus of elasticity of the string structure as shown in Eq. (\ref{equation:2}). As mentioned earlier, $E_s$  is Young’s modulus of the cell wall material, $A$ is the cross-sectional area of cell walls (i.e., for a circular cross-section with unit depth, $A = \pi r^2/D_s$), $I$ is the second moment of area of the wall’s cross-section (cell walls are assuming circular cross-section with radius $r$, and considering unit depth, i.e., $I =\frac{\pi}{64}(D_s)^3/D_s$, where $D_s=2r$ is the diameter of the string.

\begin{equation}
U=\frac{2{(Fcos\beta+Psin\beta)}^2\frac{L}{2}}{2E_sA}
  +2\int_{0}^{\frac{L}{2}}\frac{(x{(Fsin\beta-Pcos\beta)})^2}{2E_sI} \,dx
\label{equation:1}
\end{equation}

\begin{equation}
E_{sx} = \frac{12n_uwAI}{12ILcos^2\beta+L^3Asin^2\beta}\times{E_s}
\label{equation:2}
\end{equation}

Now, for finding the displacement in the transverse direction, we apply $\delta_y=\frac{\delta{u}}{\delta{P}}\textbar{_{P=0}}$, which gives $\delta{_y} = \frac{12FLIcos\beta sin\beta-FL^3Acos\beta sin\beta}{12E_sAI}$. The Poisson ratio of the string stucutre is negative ratio of average displacement in $y$-direction to the average strain in $x$-direction, that is $\nu{_{sx}} = -\epsilon{_y}/\epsilon{_x} = -\delta{_y}/\delta{_x}$. Then we achieved the following Eq. (\ref{equation:3})

\begin{equation}
\nu{_{sx}} = \frac{L^3Acos\beta sin\beta-12LIcos\beta sin\beta}{12ILcos^2\beta+L^3Asin^2\beta}
\label{equation:3}
\end{equation}

Following that, the modulus of elasticity of the string is later weighted with the volume of the string and substrate to find out the $\nu_{x}$ of the string-substrate system in $x$-direction as shown  in Eq. (\ref{equation:14}). Similarly, to determine the overall Poisson ratio of the structure, we employed a volumetric weighted ratio as shown in Eq. (\ref{equation:13}) that combines the Poisson's ratios of the individual components to calculate the overall Poisson's ratio of the structure. Here, the volume of the substrate is $V_b=L_bwh$. Taking into account the string structure as equivalent to the plate structure length and width along with string thickness $2r$, the equivalent volume of the string structure is $V_s=L_bw(2r$). Where $L_b$, is the length of the substrate. 

\begin{equation}
E_x = \frac{V_bE_b+V_sE_{sx}}{V_b+V_s}
\label{equation:14}
\end{equation}

\begin{equation}
\nu_x = \frac{V_b\nu_b+V_s\nu_{sx}}{V_b+V_s}
\label{equation:13}
\end{equation}

After finding these material properties, we obtain the Eq. (\ref{equation:11}) for moment $M_x$ using Kirchhoff plate theory. In Eq. (\ref{equation:11}), $C_f$, $D_x = \frac{wE_xh^3}{12(1-\nu_x^2)}$, $\nu_x$, $k$, and $\tau$ are correction factor, combine flexural rigidity in $x$-direction, Poisson ratio in $x$-direction, curvature in $x$-direction and curvature in $y$-direction respectively.

\begin{equation}
M_x = C_fD_x(\kappa+\nu_x\tau)
\label{equation:11}
\end{equation}

\section{Stress in $Y$-direction}
Similar to the $x$-direction, we applied Castiglianos theorem  in Eq. (\ref{equation:4}), that gives the displacement in axial direction $\delta{_y} =\frac{\delta{u}}{\delta{F}}\textbar{_{P=0}}= \frac{FL}{E_sA}(sin^2\beta)+\frac{FL^3}{12E_sI}(cos^2\beta)$. Next, we calculated the average strain $\epsilon{_y}$ of the structure in the $y$-direction, where $\epsilon{_y} = \frac{\delta{_y}}{Lsin\beta}$. The ratio of axial displacement $\sigma{_y}$ to the average strain ($\epsilon{_y}$) gives us the effective Young's modulus ($E_{yu}$) of the structure for a unit cell, $E_{yu} = \frac{12AI}{12ILsin^2\beta+L^3Acos^2\beta}\times{E_s}$.  Finally, we get Eq. (\ref{equation:5}), by multiplying with number of unit and width of the structure, where $n_u$ total number of unit cells in the system and $E_{sy}$ is the structure young's modulus in $y$-direction.

\begin{equation}
U=\frac{2{(Fsin\beta+Pcos\beta)}^2\frac{L}{2}}{2E_sA}
  +2\int_{0}^{\frac{L}{2}}\frac{(y{(Fsin\beta-Pcos\beta)})^2}{2E_sI} \,dy\\
\label{equation:4}
\end{equation}


\begin{equation}
E_{sy} = \frac{12n_uwAI}{12ILsin^2\beta+L^3Acos^2\beta}\times{E_s}
\label{equation:5}
\end{equation}

Applying, $\delta_x=\frac{\delta{u}}{\delta{P}}\textbar{_{P=0}}$ provides the displacement in transverse direction, $\delta{_x} = \frac{12FLIcos\beta sin\beta-FL^3Acos\beta sin\beta}{12E_sAI}$. 
We determined Poisson ratio by applying $\nu{_{sy}} = -\epsilon{_x}/\epsilon{_y} = -\delta{_x}/\delta{_y}$ relationship, just like we did for $x$-direction, which gives us Eq. (\ref{equation:6}). 

\begin{equation}
\nu{_{sy}} = \frac{L^3Acos\beta sin\beta-12LIcos\beta sin\beta}{L^3Acos^2\beta+12LIAsin^2\beta}
\label{equation:6}
\end{equation}

Then, We applied the volumetric weighted average of substrate and string to find out the modulus of elasticity and Poisson ratio of the structure in the $y$-direction, as shown in Eq. (\ref{equation:15}) and Eq. (\ref{equation:16}).

\begin{equation}
E_y = \frac{V_bE_b+V_sE_{sy}}{V_b+V_s}
\label{equation:15}
\end{equation}

\begin{equation}
\nu_y = \frac{V_b\nu_b+V_s\nu_{sx}}{V_b+V_s}
\label{equation:16}
\end{equation}

After identifying these material properties, we utilized Kirchhoff plate theory to develop the desired moment Eq. (\ref{equation:12}). In this equation  $C_f$, $D_y = \frac{wE_yh^3}{12(1-\nu_y^2)}$, $\nu_y$, $k$,  and $\tau$ are correction factor, combine flexural rigidity in $y$-direction, combined Poisson ratio in $y$-direction, curvature in $x$-direction and curvature in $y$-direction respectively.
\begin{equation}
M_y = C_fD_y(\tau+\nu_y\kappa)
\label{equation:12}
\end{equation}

For the purpose of determining the correction factor $C_f$ in the moment equation, we simulate the FE design without the string in two different scenarios: (1) the simulation with the substrate only, and (2) the simulation with the substrate and button. By comparing the results, we achieved the correction factor $C_f$ = 1.05.




\vspace{5mm}
\nocite{*}